\documentclass[prl,twocolumn,superscriptaddress,amsmath,amssymb,showpacs]{revtex4}
\usepackage{fancyhdr}
\usepackage{graphicx}
\usepackage{amsmath}
\usepackage{bm}
\usepackage{subfigure}
\usepackage{lastpage}

\begin{document}

\title{Gain narrowing in few-atom systems}

\author{Tom \surname{Savels}}
\email{t.savels@amolf.nl} \affiliation{FOM Institute for Atomic and
Molecular Physics, Kruislaan 407, 1098 SJ, Amsterdam, The
Netherlands}
\author{Allard P. \surname{Mosk}}
\affiliation{Complex Photonic Systems, MESA+ Research Institute,
University of Twente, PO Box 217, 7500 AE Enschede, The
Netherlands.}
\author{Ad \surname{Lagendijk}}
\affiliation{FOM Institute for Atomic and Molecular Physics,
Kruislaan 407, 1098 SJ, Amsterdam, The Netherlands}
\affiliation{Complex Photonic Systems, MESA+ Research Institute,
University of Twente, PO Box 217, 7500 AE Enschede, The
Netherlands.}

\date{May 31st, 2006}

\begin{abstract}
Using a density matrix approach, we study the simplest systems that
display both gain and feedback: clusters of 2 to 5 atoms, one of
which is pumped. The other atoms supply feedback through multiple
scattering of light. We show that, if the atoms are in each other's
near-field, the system exhibits large gain narrowing and spectral
mode redistribution. The observed phenomena are more pronounced if
the feedback is enhanced. Our system is to our knowledge the
simplest exactly solvable microscopic system which shows the
approach to laser oscillation.
\end{abstract}

\maketitle \lfoot{} \fancyhead[LO]{\footnotesize{Gain narrowing in a
few-atom system}} \fancyhead[RO]{\footnotesize{Tom Savels, Allard P.
Mosk and Ad Lagendijk}} \cfoot{\thepage\ of \pageref{LastPage}}
\pagestyle{fancy}

In a laser, light is generated by a combination of light
amplification by stimulated emission and optical feedback
\cite{Siegman1986-Haken1984-Mandel1995}. In order to study the basic
physics of these processes, there has been an intensive search for
laser operation in fundamental systems \cite{Lamb1999}. The
resulting drive toward miniaturization has led to, among others, the
realization of vertical-cavity semiconductor lasers
\cite{Chang-Hasnain1998}, dye microsphere lasers \cite{Tzeng1984},
microring and microdisk semiconductor lasers \cite{Mohideen1994} and
photonic bandgap lasers \cite{Lee1999}. As laser systems are made
smaller, a purely macroscopic description becomes inadequate and
microscopic considerations should be taken into account. An
interesting example of lasers which require a (partially)
microscopic treatment is the class of one-atom lasers
\cite{An1994-McKeever2003}, in which the gain medium is reduced to a
fundamental level, while macroscopic mirrors provide feedback.
Another, contrasting example is the class of random lasers
\cite{Cao2001-Letokhov1968-Lawandy1994}, in which optical feedback
is provided by scattering from microscopic particles, while the gain
medium remains macroscopic. Obviously, neither the feedback
mechanism nor the gain
medium can be reduced to less than one atom.\\
\indent In this Letter, we explore the most fundamental system
displaying both gain and feedback: a single pumped atom, surrounded
by one or more passive atoms providing optical feedback by
scattering. The atoms are positioned in free space in the absence of
a cavity. The absence of a cavity and its modes differentiates our
model from models in which atoms interact via a single field mode
such as, e.g., atoms in a single-atom maser \cite{Meschede1985}. Our
few-atom system can be described fully microscopically, without any
quasi-classical or paraxial approximations. We show that this
system, though very simple, shows surprisingly strong spectral gain
narrowing and mode redistribution, indicating an approach to laser
oscillation. We further demonstrate that the observed phenomena are
more pronounced as the number of atoms increases, in correspondence
with the
intuitive $N\rightarrow\infty$ limit.\\
\indent The ``atoms'' could be implemented as any type of
sub-wavelength quantum objects, for example: trapped atoms, quantum
dots \cite{Gammon2002}, trapped ions \cite{Leibfried2003} or dye
molecules. Each atom interacts with the electromagnetic field by its
transition dipole moment, which results in scattering of light. One
of the atoms is continuously pumped, causing it to not only scatter,
but also amplify the light. This $N$-atom system has optical
feedback due to the fact that in the process of stimulated emission,
a single atom scatters the stimulated photon isotropically, in
contrast to the general notion that stimulated emission preserves
the ``direction'' of the photon, which is only
true in a macroscopic gain medium \cite{Scully1997}.\\
\indent We assume the atoms to be fixed in space, e.g., by a tight
trap or a solid matrix. Each atom has three relevant energy levels:
the ground state $a$, a highly excited state $b$ and the upper state
of the relevant $c\rightarrow a$ transition $c$, as depicted in
Figure \ref{fig,threelevels} (generalization to a four-level system
is straightforward). One of the atoms is pumped with light which is
resonant with the $b$ $\rightarrow$ $a$ transition. The pump
intensity is expressed by the dimensionless parameter
$W=\Omega^2/\Gamma_{bc}\Gamma_{ca}$ , where $\Omega=|{\bm
d}_{ba}\cdot{\bm E}_{pump}|/\hbar$ is the Rabi frequency of the pump
field, with ${\bm d}_{ba}$ the dipole moment of the $b$
$\rightarrow$ $a$ transition \cite{Savels2005}. The decay rates from
$b$ to $c$ and from $c$ to $a$ are given by $\Gamma_{bc}$ and
$\Gamma_{ca}$ respectively. We can consider the atoms to be
effective two-level ($a$-$c$) systems if decay from $c$ to $a$ is
much slower than from $b$ to $c$ and decay from $b$ to $a$ is
negligible compared to other decay processes, as is usually the case
in a
three-level laser.\\
\begin{figure}[t]
    \includegraphics[width=3in]{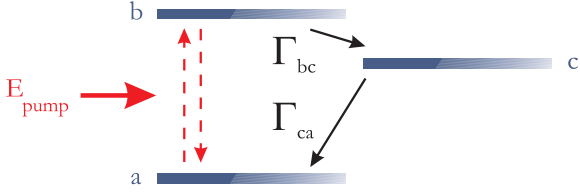}
    \caption{\label{fig,threelevels}(color online) The three-level system $a$ $b$ $c$.
    $\Gamma_{bc}$ and $\Gamma_{ca}$ are the decay rates from $b$ to $c$ and $c$ to $a$ respectively.
    Decay from $b$ to $a$ is negligible compared to other decay processes.
    The red dashed arrow expresses the interaction with the pump field.
    }
\end{figure}
\indent The ensemble-averaged populations of the atomic levels and
the coherences between them are described by a density matrix
\cite{Cohen-Tannoudji1998-Lehmberg1970-Carmichael1997}
$\hat{\sigma}(t)$, whose evolution is given by the master equation
\begin{align}
\frac{d}{dt}\hat{\sigma}&\equiv\hat{\mathcal{L}}_{nd}\hat{\sigma}+\hat{\mathcal{L}}_{d}\hat{\sigma}\label{Liouvillian},
\end{align}
with as non-dissipative operator
\begin{align}
\hat{\mathcal{L}}_{nd}\hat{\sigma}\equiv-i\sum_{i,j=1}^{N}[\delta^{ij}\hat{S}^{i+}\hat{S}^{j-},\hat{\sigma}]\label{coherent},
\end{align}
and as dissipative operator
\begin{align}
\hat{\mathcal{L}}_{d}\hat{\sigma}\equiv
&-\frac{1}{2}\sum_{i,j=1}^{N}\Gamma^{ij}\Bigl(\hat{\sigma}\hat{S}^{i+}\hat{S}^{j-}-\hat{S}^{i-}\hat{\sigma}\hat{S}^{j+}\Bigr)\nonumber\\
&-\frac{W}{2}\Gamma_{ca}\Bigl(\hat{\sigma}\hat{S}^{1-}\hat{S}^{1+}-\hat{S}^{1+}\hat{\sigma}\hat{S}^{1-}\Bigr)+\text{H.c.}.\label{incoherent}
\end{align}
The pumped atom is labeled ``1'' in expression (\ref{incoherent}).
The operators $\hat{S}^{i+}$ and $\hat{S}^{i-}$ respectively raise
and lower the state of atom $i$. The coupling between different
atoms is quantified by
\begin{align}
\delta^{mn}-\frac{i}{2}\Gamma^{mn}=3\pi\frac{\Gamma_{ca}c}{\omega_{ca}}{\bm
\mu_{m}}\cdot\mathcal{G}_{0}(\omega_{ca}, {\bm r}_{m}-{\bm
r}_{n})\cdot{\bm \mu_{n}},\label{connection}
\end{align}
for the off-diagonal elements $m\neq n$. The diagonal elements of
the coupling are given by $\Gamma^{nn}\equiv\Gamma_{ca},\forall n$
and $\delta^{nn}\equiv\omega_{ca}\equiv\omega_{c}-\omega_{a},\forall
n$. The tensor $\mathcal{G}_{0}$ represents the free-space dyadic
Green function \cite{Lagendijk1996}, $c$ is the free-space speed of
light, ${\bm r}_{i}$ the position vector of atom $i$ and ${\bm
\mu_{i}}$ is the normalized transition dipole moment of atom $i$.
The master equation (\ref{Liouvillian}) is derived by an integration
over the multimode electromagnetic field to which the atoms couple.
This integration
results in the manifestation of the effective coupling term (\ref{connection}).\\
\indent Solving the master equation (\ref{Liouvillian}) requires the
inversion and diagonalization of the associated $2^{2N}\times2^{2N}$
matrix. We solve the computer-generated symbolic master equation of
the total atomic system to calculate the spectral distribution of
the emitted light \cite{SymbolNote}. By applying a master equation,
we ensure that both elastic and inelastic scattering of photons is
taken into account in all scattering orders. Herein lies one of the
major benefits of the method we use: the atomic saturation due to
inelastically scattered photons is significant \cite{Zoller1979} and
very difficult to incorporate in a classical scattering formalism
\cite{Lagendijk1996}.\\
\indent The spectral distribution of the emitted light depends on
the atoms' spatial configuration and the orientation of the
transition dipole moments. We focus on configurations of atoms with
interatomic distances $L$ of the order $c/\omega_{ca}$. For much
larger distances, the feedback provided by the passive atoms is
limited and only a very small fraction of the photons emitted by the
pumped atom will be scattered. On the other hand, for distances much
smaller than the resonance wavelength, the photons emitted by the
pumped atom will significantly saturate the passive atoms and
effectively reduce the system's feedback. Consequently, there is a
distance range of the order of the resonance wavelength for which
the feedback provided by the passive atoms is optimal.\\
\indent The average atom-photon interaction time is of the order
$\Gamma_{ca}^{-1}$, while the time it takes for photons to propagate
from one atom to another is of the order
$\omega_{ca}^{-1}\ll\Gamma_{ca}^{-1}$. Hence, the information and
energy in the system will be stored as atomic excitations rather
than electromagnetic excitations.
The system's storage capacity is thus determined by the number of atoms.\\
\indent The spectral information of the emitted light can be deduced
from the Fourier transform of the field-correlation function
\begin{align}
g^{(1)}(\tau)\equiv\left<:\hat{\bm E}^-(t+\tau)\cdot\hat{\bm
E}^{+}(t):\right>
\end{align}
in steady-state, where the colons denote normal and time ordering
for the field operators. As a general result of the master equation,
the spectrum of the emitted light can be expressed as a sum of
$\tfrac{(2N)!}{(N+1)!(N-1)!}$ Lorentzian contributions
\cite{Steudel1979}. Each of these is characterized by a central
frequency, a spectral width and a spectral weight. The latter
expresses to what extent each contribution dominates the total
spectrum. If we increase the pump intensity, different modes will be
subject to different gain and, consequently, modes will compete for
the available population
inversion in the system.\\
\begin{figure}[t]
    \includegraphics[width=3in]{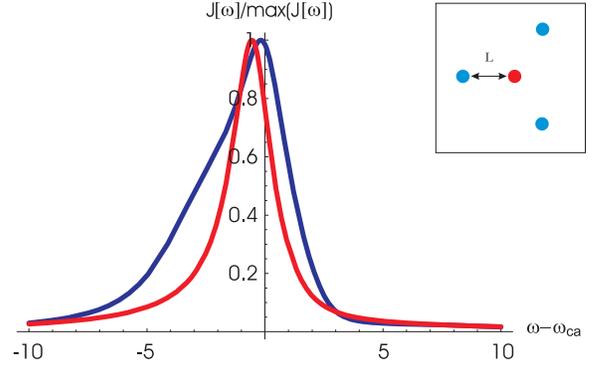}
    \caption{\label{fig,spectrum}(color) The normalized far-field angle-averaged spectrum versus frequency
    (in units of $\Gamma_{ca}$). The inset shows the configuration for which the spectrum is evaluated.
    The passive atoms (blue) are positioned in an equilateral triangle, with the pumped atom (red) in the center.
    All transition dipole moments are perpendicular to the plane
    of the atoms. The distance $L$ was chosen $0.7c/\omega_{ca}$. The spectrum is shown for
    a low pump intensity $W=1.77$ (blue) and a high pump intensity $W=10.10$
    (red).
    }
\end{figure}
\indent To indicate the effect of an increasing pump intensity on
the spectrum of the emitted light, Figure \ref{fig,spectrum} shows
the far-field spectrum for a four-atom configuration. The spectrum
is averaged over a $4\pi$ solid angle for a typical low ($W=1.77$)
and a high ($W=10.10$) pump intensity. At low pump intensity, the
emission spectrum is broad, while at higher pump intensities we
observe a significant spectral narrowing. \\
\indent If we wish to compare different configurations of atoms, we
need to visualize the degree of observed gain narrowing. We proceed
along the path originally considered by Shawlow and Townes
\cite{Schawlow1958} and compare the line width of the emitted light
to the photon emission rate. We therefore determine the full width
at half maximum $\Delta\omega$ of the far-field spectrum $I(\omega)$
averaged over a $4\pi$ solid angle. We then evaluate the spectral
weight $\int_{\Delta\omega}I(\omega)d\omega$ within the range
$\Delta\omega$, which yields the total photon emission rate
$n\Gamma_{ca}$ emitted in the range $\Delta\omega$. The number $n$
can then be interpreted as the number of excitations in a cavity
with decay rate $\Gamma_{ca}$. Since the passive atoms can store one
excitation each, the average rate $n\Gamma_{ca}$ detected in the
far-field cannot exceed $N\Gamma_{ca}$. Additionally, the average
photon emission rate is limited by the rate
$W\Gamma_{ca}\left<\hat{S}^{1-}\hat{S}^{1+}\right>$ at which pump
photons are absorbed by the system. Figure
\ref{fig,narrowing_and_inset} is a parametric plot showing the
resulting spectral width $\Delta\omega$ versus $1/n$ for the same
parameters as in Figure \ref{fig,spectrum} for increasing pump
intensity. We observe a large decrease of $\Delta\omega$,
accompanied by an increase in $n$. Since $n$ cannot increase
indefinitely, there is a critical pump intensity at which saturation
of the passive atoms sets in, as shown in the inset of Figure
\ref{fig,narrowing_and_inset}. The maximum value of $n$ is
relatively low compared to $N$ due to the weak coupling between the
atoms. Around the saturation point, the spectrum broadens while $n$
remains locally constant. If the pump intensity increases beyond the
saturation point of the passive atoms, the emission rate decreases.
This effect is due to power broadening, inherent to the three-level
pumping scheme: at the saturation point of the passive atoms, an
increase of the pump leads to a decoupling of the pumped atom and
the passive ones, resulting
in the observed decrease of $n$.\\
\indent The observed dependence of $1/n$ on $\Delta\omega$ below
saturation is similar to the behavior found in many macroscopic
lasers. As was shown by Shawlow and Townes and generalized by many
others \cite{Kuppens1994-Wiseman1999}, the quantum-limited laser
line width due to diffusion is inversely proportional to the number
of photons in the laser mode. It is striking that our simple
microscopic system exhibits a similar behavior
while we are not in the regime typically considered in the Schawlow-Townes relation.\\
\begin{figure}[t]
    \includegraphics[width=3.5in]{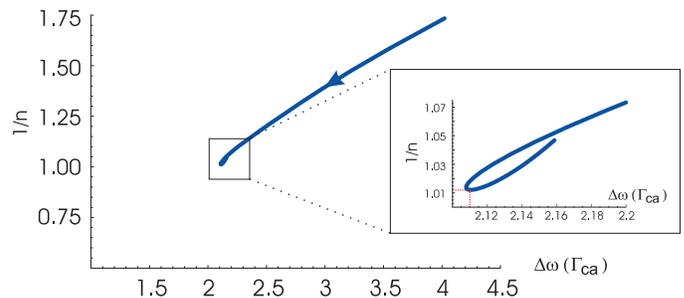}
    \caption{\label{fig,narrowing_and_inset}(color online)
    The photon emission rate (relative to $\Gamma_{ca}$) in the range $\Delta\omega$ versus
    $\Delta\omega$ (in units of $\Gamma_{ca}$). The pump intensity $W$ ranges from $1.76$ to $13.43$, while the inset
    focusses on the behavior around the saturation point. The arrow denotes an
    increase of the pump. The same configuration as in Figure \ref{fig,spectrum} is used, with $L=0.7c/\omega_{ca}$.
    The red dashed lines in the inset denote the position of $\Delta\omega_{min}$ and $n_{max}^{-1}$.
    }
\end{figure}
\indent If the number of atoms increases, the system's storage
capacity grows. Hence, the maximum value $n_{max}$ which $n$ can
attain increases with $N$. For each given number of atoms, there is
an infinite number of possible configurations in which the atoms can
be positioned. Since the dipole-dipole coupling depends on the
configuration, both $n_{max}$ and the corresponding width
$\Delta\omega_{min}$ will, for a
given $N$, vary with the geometry.\\
\indent In order to compare different configurations of atoms, we
determine how many excitations can be stored with a given coherence
time $\Delta\omega_{min}^{-1}$. For each number of atoms $N$, we
consider those configurations which attain their saturation point at
a given value of $\Delta\omega_{min}$. We then determine the
corresponding number of excitations $n_{max}$. Figure
\ref{fig,n_vs_N} shows the calculated $n_{max}$ for three different
values of the coherence time. For every number of atoms, different
configurations exist which yield the same saturation value
$\Delta\omega_{min}$. In general, such configurations each have a
different $n_{max}$ associated with them, as represented by the
identically colored symbols. We see that the effect of an increase
in $N$ is twofold. First, we observe that, for a fixed coherence
time, the maximum number of excitations increases with $N$. This
trend indicates that, as the storage capacity of the system grows,
more photons with a given coherence time $\Delta\omega_{min}^{-1}$
can be emitted by the system. Second, when comparing different
values of $\Delta\omega_{min}$ in Figure \ref{fig,n_vs_N}, we see
that, if the required coherence time increases, a larger capacity is
needed to attain a given number of excitations $n_{max}$. This
relation between the number of excitations and the storage capacity
is in accordance with the intuitive limiting case
$N\rightarrow\infty$,
$1/n_{max}\rightarrow 0$ and $\Delta\omega_{min}\rightarrow 0$.\\
\indent The efficiency with which incident pump photons are
converted into photons in the range $\Delta\omega_{min}$ is given by
the ratio of the output rate $n_{max}\Gamma_{ca}$ and the input rate
$W\Gamma_{ca}\left<\hat{S}^{1-}\hat{S}^{1+}\right>$. The numerical
value of the efficiency depends on the number of atoms and the
configuration, but we find as a general trend that the efficiency
increases with $N$. For
$\Delta\omega_{min}^{-1}=0.43\Gamma_{ca}^{-1}$ considered in Figure
\ref{fig,n_vs_N}, for example, the efficiency increases from 20\%
for $N=2$ to 24\% for $N=5$. This indicates that adding more atoms
leads to a better photon confinement, as we expect.\\
\indent The mechanism governing light amplification in the few-atom
systems presented here is stimulated emission. The spontaneous
radiation emitted by the gain atom is scattered by the passive
atoms; subsequent interaction with the gain atom generates
stimulated emission. The presented physical processes correspond to
the behavior expected of a sub-threshold
bad-cavity laser \cite{Kuppens1994-Wiseman1999,Siegman1998}.\\
\begin{figure}[t]
    \includegraphics[width=3in]{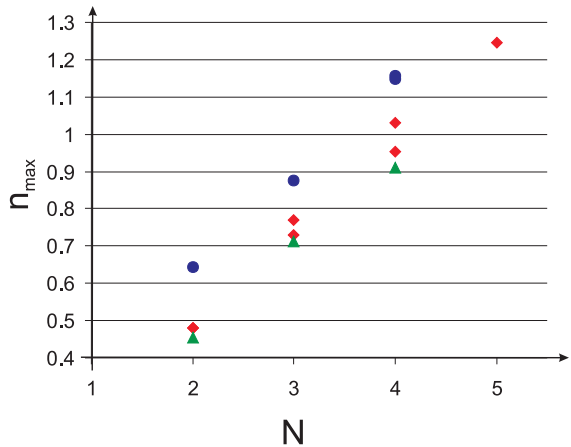}
    \caption{\label{fig,n_vs_N}(color) The relation between the maximum
    excitations number $n_{max}$ and the number of atoms $N$. Multiple symbols for a fixed $N$ represent
    different configurations. Fixed colors represent fixed coherence times.
    The green triangles are for a coherence time $0.47\Gamma_{ca}^{-1}$.
    The red diamonds are for a coherence time
    $0.43\Gamma_{ca}^{-1}$. The blue circles are for a coherence time
    $0.29\Gamma_{ca}^{-1}$.
    }
\end{figure}
\indent From an experimental point of view, we are convinced that
our model is relevant for a wide range of realizations, such as
laser cooled trapped ions \cite{Eschner2001}. Selective excitation
of the ions can be achieved by tuning the polarization of the pump
field and the positioning of the ions. Another possible experimental
path would be to implement the atoms as quantum dots
\cite{Gerardot2005} which form, if bound to DNA, bioconjugated
superstructures \cite{Lee2005-Fu2004}. While the energy transfer in
those structures is somewhat different from the one presented here,
we anticipate the gain narrowing phenomenon as presented in this
Letter to remain conceptually valid. As a third experimental
realization, we expect our results to stimulate experimental work on
cold atoms \cite{Sigwarth2004} interacting with pumping fields.
Compared to our current system, the number of atoms in a cold-atom
cloud is very
high; we therefore trust the phenomena described above to be much more pronounced.\\
\indent In conclusion, we described the simplest microscopic system
which shows both gain and feedback. A system of only a few atoms in
each other's near-field shows large gain narrowing and mode
redistribution. Surprisingly, our system qualitatively exhibits
behavior similar to macroscopic lasers. Adding more atoms to the
system enhances the observed phenomena and allows more photons to
propagate with a given
coherence time.\\
\indent Various extensions to our model are possible, among which
schemes where more than one atom is pumped. The challenge of our
method lies within the exponential scaling of the matrices involved;
other approaches such as stochastical wave function calculations
\cite{Molmer1993} might help overcome the current limits.
Conversely, our current system may serve as a building-block for
few-body contributions in an effective medium approach to, e.g.,
coherent backscattering \cite{Lagendijk1996}. Importantly, we expect
our results can be tested experimentally, and we are working towards
a physical realization of our model.
\begin{acknowledgements}
The authors wish to thank Gerhard Rempe and Bart van Tiggelen for
fruitful discussions. This work is part of the research program of
the `Stichting voor Fundamenteel Onderzoek der Materie' (FOM), which
is financially supported by the `Nederlandse Organisatie voor
Wetenschappelijk Onderzoek' (NWO).
\end{acknowledgements}


\begin{thebibliography}{99'}
\bibitem{Siegman1986-Haken1984-Mandel1995}
A.E. Siegman, \textit{Lasers} (University Science Books, Mill
Valley, 1986); H. Haken, \textit{Laser Theory} (Springer,
Berlin,1984); L. Mandel and E. Wolf, \textit{Optical Coherence and
Quantum Optics} (Cambridge Univ. Press, Cambridge, 1995)
\bibitem{Lamb1999}
W.E. Lamb \textit{et al.}, Rev. Mod. Phys. \textbf{71}, S263 (1999)
\bibitem{Chang-Hasnain1998}
C.J. Chang-Hasnain, Opt. Photonics News \textbf{9}, 35 (1998)
\bibitem{Tzeng1984}
H.-M. Tzeng \textit{et al.}, Opt. Lett. \textbf{9}, 499 (1984)
\bibitem{Mohideen1994}
U. Mohideen \textit{et al.}, Appl. Phys. Lett \textbf{64}, 1911
(1994)
\bibitem{Lee1999}
R.K. Lee \textit{et al.}, Electron. Lett. \textbf{35}, 569 (1999)
\bibitem{An1994-McKeever2003}
K. An \textit{et al.}, Phys. Rev. Lett. \textbf{73}, 3375 (1994); J.
McKeever \textit{et al.} Nature \textbf{425}, 268 (2003)
\bibitem{Cao2001-Letokhov1968-Lawandy1994}
H. Cao \textit{et al.}, Phys. Rev. Lett. \textbf{86}, 4524 (2001);
V. S. Letokhov, Sov. Phys. JETP \textbf{26}, 835 (1968); N.M.
Lawandy \textit{et al.}, Nature \textbf{368}, 436 (1994)
\bibitem{Meschede1985}
D. Meschede, H. Walther and G. M{\"u}ller, Phys. Rev. Lett.
\textbf{54}, 551 (1985)
\bibitem{Gammon2002}
D. Gammon and D.G. Steel, Phys. Today \textbf{55}, 36 (2002)
\bibitem{Leibfried2003}
D. Leibfried \textit{et al.}, Rev. Mod. Phys. \textbf{75}, 281
(2003)
\bibitem{Scully1997}
M.O. Scully and M.S. Zubairy, \textit{Quantum Optics} (Cambridge
University Press, Cambridge, 1997), Chapter 10
\bibitem{Savels2005}
T. Savels, A.P. Mosk and A. Lagendijk, Phys. Rev. A \textbf{71},
043814 (2005)
\bibitem{Cohen-Tannoudji1998-Lehmberg1970-Carmichael1997}
C. Cohen-Tannoudji, J. Dupont-Roc and G. Grynberg,
\textit{Atom-Photon Interactions} (Wiley Science Paper Series, New
York, 1998); R.H. Lehmberg, Phys. Rev. A \textbf{2}, 883 (1970)
\bibitem{Lagendijk1996}
A. Lagendijk and B.A. van Tiggelen, Physics Rep. \textbf{70}, 145
(1996)
\bibitem{SymbolNote}
This tedious calculation involves symbolically evaluating Eq.
(\ref{Liouvillian}) and subsequent numerical inversion and
diagonalization of the generated matrix for each configuration.
\bibitem{Zoller1979}
P. Zoller, Phys. Rev. A \textbf{20}, 2420 (1979)
\bibitem{Steudel1979}
H. Steudel, J. Phys. B \textbf{12}, 3309 (1979)
\bibitem{Schawlow1958}
A.L. Schawlow and C.H. Townes, Phys. Rev. \textbf{112}, 1940 (1958)
\bibitem{Kuppens1994-Wiseman1999}
S.J.M. Kuppens, M.P. van Exter and J.P. Woerdman, Phys. Rev. Lett.
\textbf{72}, 3815 (1994); H.M. Wiseman, Phys. Rev. A \textbf{60},
4083 (1999)
\bibitem{Eschner2001}
J. Eschner \textit{et al.}, Nature \textbf{413}, 495 (2001)
\bibitem{Gerardot2005}
B.D. Gerardot \textit{et al.}, Phys. Rev. Lett. \textbf{95}, 137403
(2005)
\bibitem{Lee2005-Fu2004}
J. Lee, A.O. Govorov and N.A. Kotov, Nanoletters \textbf{5}, 2063
(2005); A. Fu \textit{et al.}, J. Am. Chem. Soc. \textbf{126}, 10832
(2004)
\bibitem{Sigwarth2004}
O. Sigwarth \textit{et al.}, Phys. Rev. Lett. \textbf{93}, 143906
(2004)
\bibitem{Siegman1998}
A.E. Siegman, Phys. Today, January 1998
\bibitem{Molmer1993}
K. Molmer, Y. Castin and J. Dalibard, J. Opt. Soc. Am. B
\textbf{10}, 524 (1993)
\end{thebibliography}
\end{document}